\documentclass[10pt,floatfix,letterpaper,amsmath,amsfonts,amssymb,aps,rmp,reprint,superscriptaddress]{revtex4-1}
\usepackage[latin1]{inputenc}
\usepackage{bm}
\usepackage{verbatim}
\usepackage{graphicx}
\usepackage{tensor}
\usepackage{epstopdf}
\newcommand{\op}[1]{\hat{#1}}
\newcommand{\ket}[1]{\lvert #1\rangle}
\newcommand{\bra}[1]{\langle #1 \rvert}
\newcommand{\abs}[1]{\lvert #1 \rvert}
\newcommand{\pr}[1]{\ket{#1}\bra{#1}}
\newcommand{\braket}[2]{\langle #1 \vert #2 \rangle}
\newcommand{\mean}[1]{\left\langle #1 \right\rangle}

\begin{document}
\title{\textit{Colloquium}: Understanding Quantum Weak Values: Basics and Applications}
\author{Justin Dressel}
\affiliation{Department of Physics and Astronomy and Center for Coherence and Quantum Optics, University of Rochester, Rochester, New York 14627, USA}
\affiliation{Department of Electrical Engineering, University of California, Riverside, California 92521, USA}
\author{Mehul Malik}
\affiliation{The Institute of Optics, University of Rochester, Rochester, New York 14627, USA}
\affiliation{Institute for Quantum Optics and Quantum Information (IQOQI), Austrian Academy of Sciences, Boltzmanngasse 3, A-1090 Vienna, Austria}
\author{Filippo M. Miatto}
\affiliation{Department of Physics, University of Ottawa, Ottawa, Ontario, Canada}
\author{Andrew N. Jordan}
\affiliation{Department of Physics and Astronomy and Center for Coherence and Quantum Optics, University of Rochester, Rochester, New York 14627, USA}
\author{Robert W. Boyd}
\affiliation{The Institute of Optics, University of Rochester, Rochester, New York 14627, USA}
\affiliation{Department of Physics, University of Ottawa, Ottawa, Ontario, Canada}

\date{\today}

\begin{abstract}
Since its introduction 25 years ago, the quantum weak value has gradually transitioned from a theoretical curiosity to a practical laboratory tool. While its utility is apparent in the recent explosion of weak value experiments, its interpretation has historically been a subject of confusion. Here, a pragmatic introduction to the weak value in terms of measurable quantities is presented, along with an explanation of how it can be determined in the laboratory. Further, its application to three distinct experimental techniques is reviewed. First, as a large interaction parameter it can amplify small signals above technical background noise. Second, as a measurable complex value it enables novel techniques for direct quantum state and geometric phase determination. Third, as a conditioned average of generalized observable eigenvalues it provides a measurable window into nonclassical features of quantum mechanics. In this selective review, a single experimental configuration is used to discuss and clarify each of these applications.
\end{abstract}

\maketitle
\tableofcontents

\section{Introduction}
Derived in 1988 by Aharonov, Albert, and Vaidman \cite{Aharonov1988,Duck1989,Ritchie1991} as a ``new kind of value for a quantum variable'' that appears when averaging preselected and postselected weak measurements, the quantum weak value has had an extensive and colorful theoretical history \cite{Aharonov2008,Aharonov2010,Kofman2012,Shikano2012}.  Recently, however, the weak value has stepped into a more public spotlight due to three types of experimental applications.  It is our aim in this brief and selective review to clarify these three pragmatic roles of the weak value in experiments. 

First, in its role as an evolution parameter, a large weak value can help to amplify a detector signal and enable the sensitive estimation of unknown small evolution parameters, such as beam deflection \cite{Hosten2008,Dixon2009,Starling2009,Turner2011,Hogan2011,Pfeifer2011,Zhou2012,Zhou2013,Jayaswal2014}, frequency shifts \cite{Starling2010}, phase shifts \cite{Starling2010b}, angular shifts \cite{Magana2013}, temporal shifts \cite{Brunner2010,Strubi2013}, velocity shifts \cite{Viza2013}, and even temperature shifts \cite{Egan2012}.  Paradigmatic optical experiments that have used this technique include the measurement of 1\AA\ resolution beam displacements due to the quantum spin Hall effect of light ``without the need for vibration or air-fluctuation isolation'' \cite{Hosten2008},  an angular mirror rotation of 400frad due to linear piezo motion of 14fm using only 63$\mu$W of power postselected from 3.5mW total beam power \cite{Dixon2009}, and a frequency sensitivity of 129kHz/$\sqrt{\textrm{Hz}}$ obtained with 85$\mu$W of power postselected from 2mW total beam power \cite{Starling2010}.  All these results were obtained in modest tabletop laboratory conditions, which was possible since the technique amplifies the signal above certain types of technical noise backgrounds (e.g., electronic $1/f$ noise or vibration noise) \cite{Starling2009,Feizpour2011,Jordan2014,Knee2014}.

Second, in its role as a complex number whose real and imaginary parts can both be measured, the weak value has encouraged new methods for the direct measurement of quantum states \cite{Lundeen2011,Lundeen2012,Salvail2013,Lundeen2014,Malik2014} and geometric phases \cite{Sjoqvist2006,Kobayashi2010,Kobayashi2011}.  These methods express abstract theoretical quantities such as a quantum state in terms of complex weak values, which can then be measured experimentally.  Notably, the real and imaginary components of a quantum state in a particular basis can be directly determined with minimal postprocessing using this technique.

Third, in its role as a conditioned average of generalized observable eigenvalues, the real part of the weak value has provided a measurable window into nonclassical features of quantum mechanics.  Conditioned averages outside the normal eigenvalue range have been linked to paradoxes such as Hardy's paradox \cite{Aharonov2002,Lundeen2009,Yokota2009} and the three-box paradox \cite{Resch2004}, as well as the violation of generalized Leggett-Garg inequalities that indicate nonclassical behavior \cite{Palacios-Laloy2010,Goggin2011,Dressel2011,Suzuki2012,Emary2014,Groen2013}.  Conditioned averages have also been used to experimentally measure physically meaningful quantities including superluminal group velocities in optical fiber \cite{Brunner2004}, momentum-disturbance relationships in a two-slit interferometer \cite{Mir2007}, and locally averaged momentum streamlines passing through a two-slit interferometer \cite{Kocsis2011} [i.e., along the energy-momentum tensor field \cite{Hiley2012}, or Poynting vector field \cite{Bliokh2013,Dressel2014}].

This Colloquium is structured as follows. In the next two sections we explain what a weak value is and how it appears in the theory quite generally. We then explain how it is possible to measure both its real and imaginary parts and explore the three classes of experiments outlined above that make use of weak values. This approach allows us to address the importance and utility of weak values in a clear and direct way without stumbling over interpretations that have historically tended to obscure these points.  Throughout this Colloqium, we make use of one simple notation for expressing theoretical notions, and one experimental setup --- a polarized beam passing through a birefringent crystal. 

\section{What is a weak value?}\label{sec:whatis}

First introduced by \citet{Aharonov1988}, weak values are complex numbers that one can assign to the powers of a quantum observable operator $\op{A}$ using two states: an initial state $\ket{i}$, called the \emph{preparation} or \emph{preselection}, and a final state $\ket{f}$, called the \emph{postselection}.  The $n$\textsuperscript{th} order weak value of $\op{A}$ has the form
\begin{align}\label{eq:awv}
  A^n_w &= \frac{\bra{f}\op{A}^n\ket{i}}{\braket{f}{i}},
\end{align}
where the order $n$ corresponds to the power of $\op{A}$ that appears in the expression. In this Colloquium, we clarify how these peculiar complex expressions appear naturally in laboratory measurements. To accomplish this goal, we derive them in terms of measurable detection probabilities. Weak values of every order appear when we characterize how an intermediate interaction affects these detection probabilities. 

Consider a standard prepare-and-measure experiment. If a quantum system is prepared in an initial state $\ket{i}$, the probability of detecting an event corresponding to the final state $\ket{f}$ is given by the squared modulus of their overlap $P=\abs{\braket{f}{i}}^2$. If, however, the initial state is modified by an intermediate unitary interaction $\op{U}(\epsilon)$, the detection probability also changes to $P_\epsilon=\abs{\braket{f}{i'}}^2=\abs{\bra{f}\op{U}(\epsilon)\ket{i}}^2$.

In order to calculate the relative change between the original and the modified probability, we must examine the unitary operator $\hat U(\epsilon)$ carefully. In quantum mechanics, any observable quantity is represented by a Hermitian operator. Stone's theorem states that any such Hermitian operator $\hat A$ can generate a continuous transformation along a complementary parameter $\epsilon$ via the unitary operator $\hat U(\epsilon)=\exp(-i\epsilon \hat A)$. For instance, if $\hat A$ is an angular momentum operator, the unitary transformation generates rotations through an angle $\epsilon$, or if $\hat A$ is a Hamiltonian, the unitary operator generates translations along a time interval $\epsilon$, and so on.  In \cite{Aharonov1988} (and most subsequent appearances of the weak value) $\op{A}$ is chosen to be an impulsive interaction Hamiltonian of product form; we return to this special case in Section~\ref{sec:measure}. 

If $\epsilon$ is small enough, or in other words if $\op{U}(\epsilon)$ is ``weak,'' we can consider its Taylor series expansion. The detection probability introduced above can then be written as (shown here to first order)
\begin{align}\label{eq:pepsilongen}
  P_\epsilon &= |\bra{f}\op{U}(\epsilon)\ket{i}|^2 = |\bra{f}(1-i\epsilon\hat A+\dots)\ket{i}|^2\nonumber\\
  &=P  + 2 \epsilon\, \text{Im}\braket{i}{f}\bra{f}\op{A}\ket{i} + O(\epsilon^2).
\end{align}
As long as $\ket{i}$ and $\ket{f}$ are not orthogonal (i.e. $P\neq0$), we can divide both sides of Eq. (2) by $P$ to obtain the relative correction (shown here to second order):
\begin{align}\label{eq:singlewv}
  \frac{P_\epsilon}{P} &= 1 + 2 \epsilon\, \text{Im}A_w - \epsilon^2\left[ \text{Re}A^2_w - |A_w|^2\right] + O(\epsilon^3),
\end{align}
where $A_w$ is the first order weak value and $A^2_w$ is the second order weak value as defined in Eq. (1). Here we arrive at our operational definition: weak values characterize the relative correction to a detection probability $\abs{\braket{f}{i}}^2$ due to a small intermediate perturbation $\op{U}(\epsilon)$ that results in a modified detection probability $\abs{\bra{f}\op{U}(\epsilon)\ket{i}}^2$.  Although we show the expansion only to second order here, we emphasize that the full Taylor series expansion for $P_\epsilon/P$ is completely characterized by complex weak values $A^n_w$ of all orders $n$ \cite{DiLorenzo2012,Kofman2012,Dressel2012e}. 

When the higher order terms in the expansion \eqref{eq:singlewv} can be neglected, one has a linear relationship between the probability correction and the first order weak value, which we call the \emph{weak interaction regime}.  These terms can be neglected under two conditions: (a) the relative correction $P_\epsilon / P - 1$ is itself sufficiently small, and (b) $\epsilon\text{Im}A_w$ is sufficiently large compared to the sum of higher order corrections \cite{Duck1989}. When these conditions do not hold (such as when $P\to 0$), the terms involving higher order weak values $A^n_w$ become significant and can no longer be neglected \cite{DiLorenzo2012}. Most experimental work involving weak values has been done in the weak interaction regime characterized by the first order weak value, so we will limit our discussion to that regime as well. In Section~\ref{sec:measure}, we put these ideas in the context of a real optics experiment and discuss how one measures weak values in the laboratory.

\section{How does one measure a weak value?}\label{sec:measure}

\begin{figure}[t!]
\centering
\includegraphics[width=0.9\columnwidth]{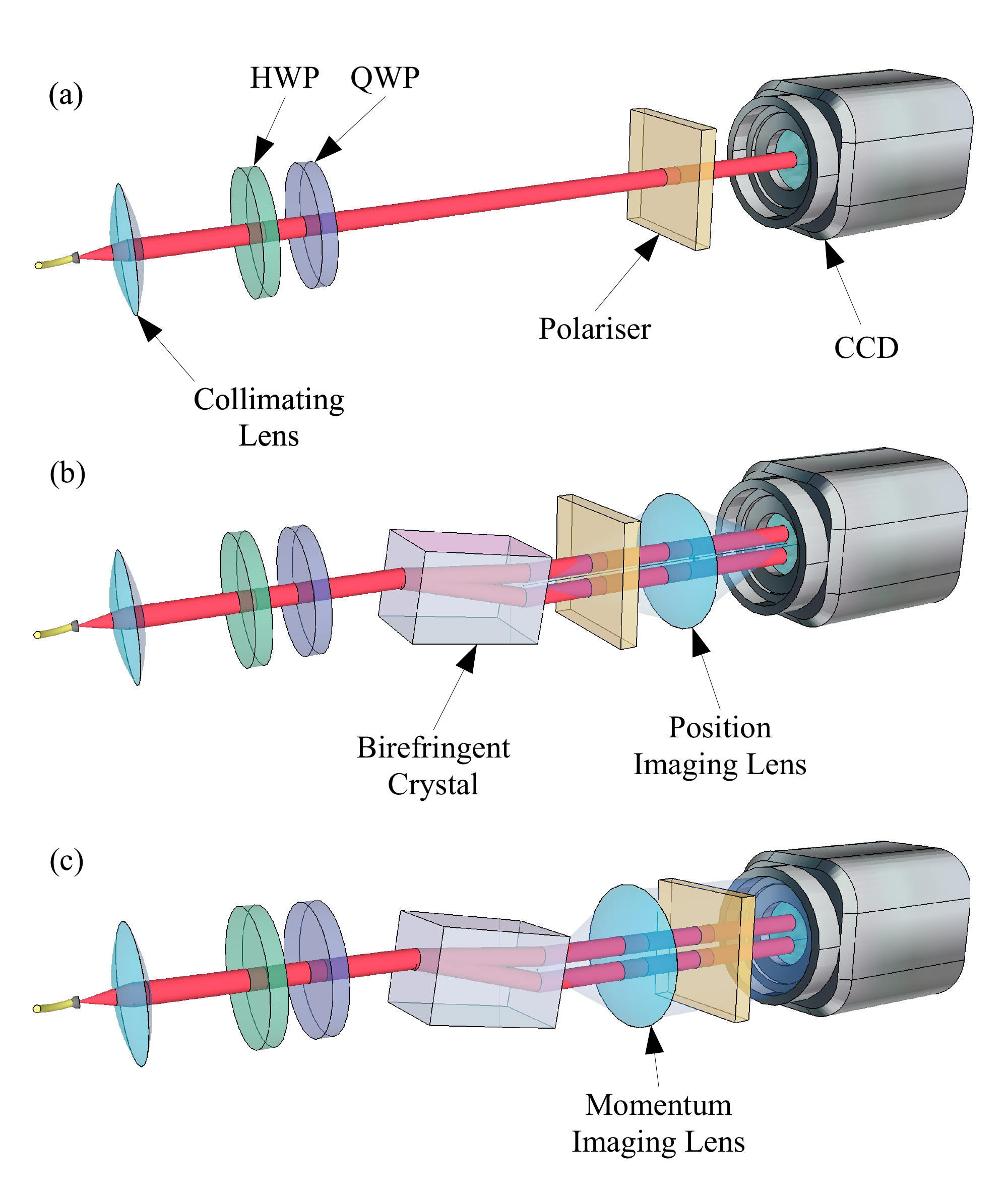}
\caption{An experiment for illustrating how one can measure weak values. (a) A Gaussian beam from a single mode fiber is collimated by a lens and prepared in an initial polarization state by a quarter-wave plate (QWP) and half-wave plate (HWP). A polarizer postselects the beam on a final polarization state. A CCD then measures the position-dependent beam intensity. (b) A birefringent crystal is inserted between the wave plates and polarizer to displace different polarizations by a small amount. A lens images the transverse position on the output face of the crystal onto the CCD in order to measure the real part of the polarization weak value as a linear shift in the postselected intensity. (c) The lens is changed to imaging the far-field of the crystal face onto the CCD as the transverse momentum in order to determine the imaginary part of the polarization weak value (details in the text).}\label{crystal}
\end{figure}

In general, weak values are complex quantities. In order to determine a weak value, one must be able to measure both its real and imaginary parts. Here, we use an optical experimental example to show how one can measure a complex weak value associated with a polarization observable. Although this particular example can also be understood using classical wave mechanics \cite{Howell2010,Brunner2003}, the quantum mechanical analysis we provide here has wider applicability.

Consider the setup shown in Fig. \ref{crystal}(a). A collimated laser beam is prepared in an initial state $\ket{i}\ket{\psi_i}$, where $\ket{i}$ is an initial polarization state and $\ket{\psi_i}$ is the state of the transverse beam profile.  The polarization is prepared through the use of a quarter-wave plate (QWP) and a half-wave plate (HWP). The beam then passes through a linear polarizer aligned to a final polarization state $\ket{f}$ before impacting a charge coupled device (CCD) image sensor for a camera. Each pixel of the CCD measures a photon of this beam with a detection probability given by 
\begin{align}\label{eq:p}
  P = \abs{\braket{f}{i}}^2\abs{\braket{\psi_f}{\psi_i}}^2,
\end{align}
where $\ket{\psi_f}$ is the final transverse state postselected by each pixel. For our purposes, this state corresponds to either a specific transverse position $\ket{\psi_f} = \ket{x}$ or transverse momentum $\ket{\psi_f}=\ket{p}$, depending on whether we image the position or the momentum space onto the CCD [e.g., using a Fourier lens as shown in Fig. \ref{crystal}(c)]. We will refer to this detection probability $P$ as the ``unperturbed'' probability. 

We now introduce a birefringent crystal between the preparation wave plates and the postselection polarizer, as shown in Fig. \ref{crystal}(b). The crystal separates the beam into two beams with horizontal and vertical polarizations. The transverse displacements depend on the birefringence properties of the crystal and on the crystal length. We assume that the crystal is tilted with respect to the incident beam so that each polarization component is displaced by an equal amount $\epsilon=\tau v$ where $\tau$ is the time spent inside the crystal and $v$ is the displacement speed.

The effect of the birefringent crystal can be expressed by a time evolution operator $\op{U}(\tau) = e^{-i\, \tau \op{H}/\hbar}$ with an effective interaction Hamiltonian
\begin{align}\label{eq:ham}
  \op{H} = v\op{S}\otimes\op{p}.
\end{align}
Here, $\op{S} = \pr{H} - \pr{V}$ is the Stokes polarization operator that assigns eigenvalues $+1$ and $-1$ to the $\ket{H}$ and $\ket{V}$ polarizations, respectively, and $\op{p}$ is the transverse momentum operator that generates translations in the transverse position $x$. This time evolution operator $\op{U}(\tau)$ correlates the polarization components of the beam with their transverse position by translating them in opposite directions. Each pixel of the CCD then collects a photon with a ``perturbed'' probability given by
\begin{align}\label{eq:pepsilon}
  P_\epsilon = |\bra{f}\bra{\psi_f}e^{-i\epsilon \op{S}\otimes\op{p}/\hbar}\ket{i}\ket{\psi_i}|^2,
\end{align}
which has the form of Eq.~\eqref{eq:pepsilongen} with the generic operator $\op{A}$ replaced by the product operator $\op{S}\otimes\op{p}$.

\begin{figure}[t]
  \begin{center}
  \includegraphics[width=0.6\columnwidth]{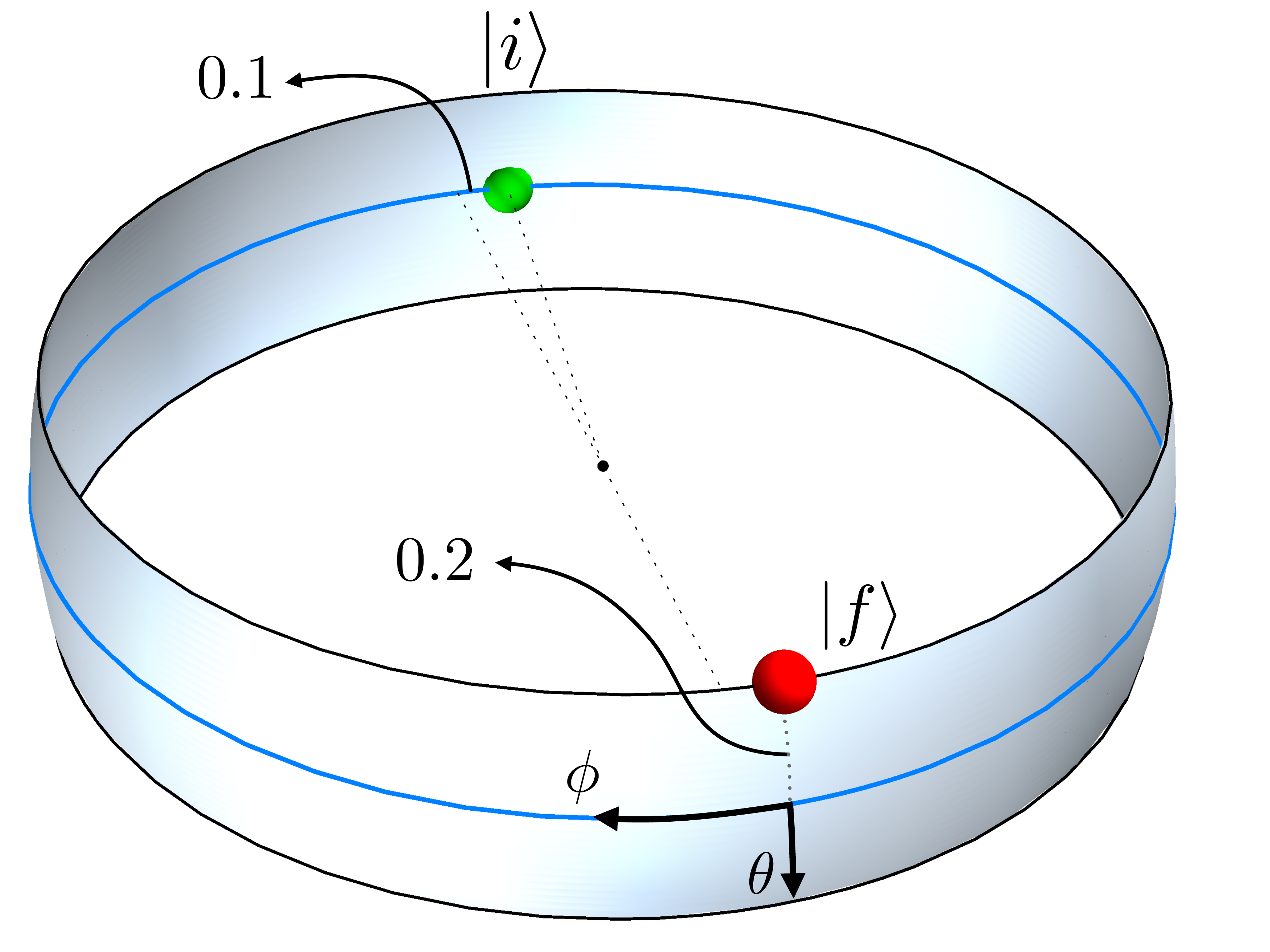}
  \end{center}
  \caption{A band around the equator of the Poincar\'e sphere showing the initial polarization $\ket{i}$ (dot, back of sphere) from Eq.~\eqref{eq:ipol} and postselection polarization $\ket{f}$ (dot, front of sphere) from Eq.~\eqref{eq:fpol}. We also indicate the small angles that make $\ket{f}$ almost orthogonal to $\ket{i}$.}
  \label{fig:gauss1}
\end{figure}

As a visual example, consider a Gaussian beam 
\begin{align}\label{eq:gaussian}
  \braket{x}{\psi_i} &= (2\pi\sigma^2)^{-1/4} \, \exp\left(-\frac{x^2}{4\sigma^2}\right),
\end{align}
with an initial antidiagonal polarization preparation with a slight ellipticity:
\begin{align}\label{eq:ipol}
  \ket{i} &= \frac{\ket{H} - e^{i\phi}\ket{V}}{\sqrt2}, \qquad \phi=0.1,
\end{align}
that passes through a linear postselection polarizer that is oriented at a small angle ($0.2$ rad in this example) from the diagonal state:
\begin{align}\label{eq:fpol}
  \ket{f} &= \cos\frac{\theta}{2}\ket{H} + \sin\frac{\theta}{2}\ket{V}, & \theta &= \frac{\pi}{2}-0.2.
\end{align}

\begin{figure}[t]
  \begin{center}
  \includegraphics[width=\columnwidth]{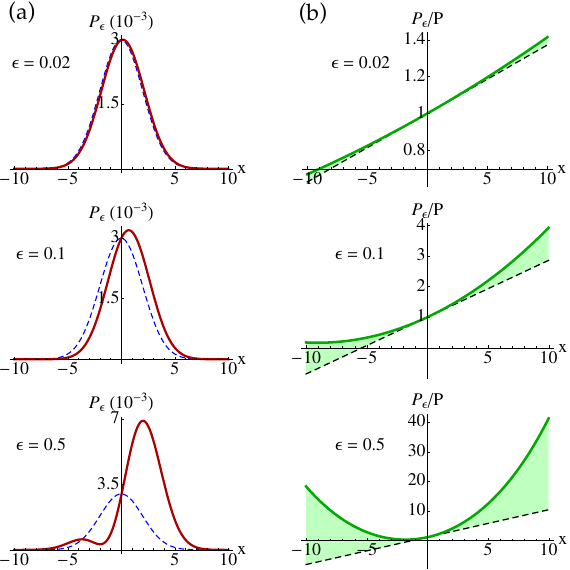}
  \end{center}
  \caption{(a) Comparisons between perturbed profiles (solid, for various values of beam displacement $\epsilon$) and a fixed unperturbed profile (dashed, corresponding to $P$). Note that both curves represent postselected measurements. (b) The exact ratio of the two curves (solid) is compared to the first order approximation (dashed). When $\epsilon$ is sufficiently small, the first order approximation adequately models the quantity $P_\epsilon/P$ over most of the profile.
  }
  \label{fig:mult1}
\end{figure}

These two nearly orthogonal polarization states are shown on a band around the equator of the Poincar\'e sphere in Fig.~\ref{fig:gauss1}.  Without the crystal present [Fig. \ref{crystal}(a)], the CCD measures the initial Gaussian intensity profile shown as a dashed line in Fig.~\ref{fig:mult1}(a) with a total postselection probability given by $\abs{\braket{f}{i}}^2=0.012$. When the crystal is present [Fig. \ref{crystal}(b)], the orthogonal polarization components become spatially separated by a displacement $\epsilon$ before passing through the postselection polarizer. The measured profiles for different crystal lengths are shown as the solid line distributions in Fig. \ref{fig:mult1}(a). The dotted line distributions show the unperturbed (but still postselected) profiles for comparison.

In the weak interaction regime, the crystal is short, $\epsilon$ is small, and the two orthogonally polarized beams are displaced by a small amount before they interfere at the postselection polarizer. As shown in Section~\ref{sec:whatis}, we can expand the ratio between the perturbed and unperturbed probabilities to first order in $\epsilon$ and isolate the linear probability correction term:
\begin{align}\label{eq:spwv}
  \frac{P_{\epsilon}}{P} - 1 &\approx \frac{2\tau}{\hbar}\, \text{Im}H_w \\
  &= \frac{2\epsilon}{\hbar}\left[\text{Re}S_w\text{Im}p_w + \text{Im}S_w\text{Re}p_w\right]. \nonumber
\end{align}
Since the Hamiltonian from Eq. \eqref{eq:ham} is of product form, its first order weak value contribution $\text{Im}H_w$ expands to a symmetric combination of the real and imaginary parts of the weak values of polarization $S_w = \bra{f}\op{S}\ket{i}/\braket{f}{i}$ and momentum $p_w = \bra{\psi_f}\op{p}\ket{\psi_i}/\braket{\psi_f}{\psi_i}$.  A clever choice of preselection and postselection states therefore allows an experimenter to isolate each of these quantities using different experimental setups \cite{Aharonov1988,Jozsa2007,Shpitalnik2008}.

To illustrate this idea for the polarization weak value, the procedure for measuring the real part $\text{Re}S_w$ is shown in Fig. \ref{crystal}(b). We image the output face of the crystal onto the CCD so that each pixel corresponds to a postselection of the transverse position $\ket{\psi_f} = \ket{x}$.  As a result, the momentum weak value for each pixel becomes 
\begin{align}\label{eq:imagpw}
  p_w &= \frac{\bra{x}\op{p}\ket{\psi_i}}{\braket{x}{\psi_i}} = \frac{-i\hbar \partial_x \psi_i(x)}{\psi_i(x)} = i\hbar \frac{x}{2\sigma^2},
\end{align}
using the Gaussian profile in Eq.~\eqref{eq:gaussian}.

Since this expression is purely imaginary, Eq.~\eqref{eq:spwv} simplifies to
\begin{align}
\label{eq:Pex}
\frac{P_\epsilon}{P} \approx 1 + \epsilon \frac{x}{\sigma^2}\, \text{Re}S_w,
\end{align}
effectively isolating the quantity $\text{Re}S_w$ to first order in $\epsilon$. The solid curves in Fig.~\ref{fig:mult1}(b) illustrate the ratio $P_\epsilon/P$ as a function of $x$ for different values of $\epsilon$. When $\epsilon$ is sufficiently small, the expansion of $P_\epsilon/P$ to first order in Eq.~\eqref{eq:Pex} [dashed lines in Fig.~\ref{fig:mult1}(b)] is a good approximation over most of the beam profile.  Pragmatically, this means that one can average the whole beam profile and still retain a linear correction that is proportional to $\text{Re}S_w$, as done originally by \citet{Aharonov1988}.

The analogous procedure for measuring the imaginary part $\text{Im}S_w$ is shown in Fig. \ref{crystal}(c). We image the Fourier plane of the crystal onto the CCD so that each pixel corresponds to a postselection of the transverse momentum $\ket{\psi_f} = \ket{p}$.  As a result, the momentum weak value for each pixel becomes simply
\begin{align}\label{eq:realpw}
  p_w &= \frac{\bra{p}\op{p}\ket{\psi_i}}{\braket{p}{\psi_i}} =\frac{p\braket{p}{\psi_i}}{\braket{p}{\psi_i}} = p.
\end{align}
Since this expression is now purely real, Eq.~\eqref{eq:spwv} simplifies to 
\begin{align}
\label{eq:Pep}
\frac{P_\epsilon}{P} \approx 1 + \epsilon\frac{2p}{\hbar}\, \text{Im}S_w,
\end{align}
effectively isolating the quantity $\text{Im}S_w$ to first order in $\epsilon$.  As with Eq.~\eqref{eq:Pex}, this first order expansion is a good approximation over most of the Fourier profile when $\epsilon$ is sufficiently small.  Hence, the profile may be similarly averaged and retain the linear correction proportional to $\text{Im}S_w$, as done originally in \citet{Aharonov1988}.

Note that we could also isolate the real and imaginary parts of $p_w$ in a similar manner through a judicious choice of polarization postselection states.  More generally, one can use this technique to isolate weak values of any desired observable by constructing Hamiltonians in a product form such as Eq.~\eqref{eq:ham} and cleverly choosing the preselection and postselection of the auxiliary degree of freedom.

\section{How can weak values be useful?}\label{sec:useful}
In Section~\ref{sec:measure}, we showed how the relative change in postselection probability can be completely described by complex weak value parameters. We also elucidated how the real and imaginary parts of the first order weak value can be isolated and therefore measured in the weak interaction regime. 

In this section we focus on three main applications of the first order weak value.  First, we show how clever choices of the initial and final postselected states can result in large weak values that can be used to sensitively determine unknown parameters affecting the state evolution.  Second, we show how the complex character of the weak value may be used to directly determine a quantum state.  Third, we show how the real part of the weak value can be interpreted as a form of conditioned average pertaining to an observable.

\subsection{Weak value amplification}
In precision metrology an experimenter is interested in estimating a small interaction parameter, such as the transverse beam displacement $\epsilon = \tau v$ due to the crystal in Section~\ref{sec:measure}. As the first order approximation of $P/P_\epsilon$ holds in the weak interaction regime, the value of $\epsilon$ can be directly determined.  We briefly note that the appearance of the joint weak value of Eq.~\eqref{eq:spwv} in a parameter estimation experiment is no accident: as pointed out by \citet{Hofmann2011b}, this quantity is the \emph{score} used to calculate the Fisher information that determines the Cramer-Rao bound for the estimation of an unknown parameter such as $\epsilon$ \cite{Helstrom1976,Hofmann2012,Viza2013,Jordan2014,Pang2014,Knee2014}.

Being able to resolve a small $\epsilon$ in the presence of background noise requires the joint weak value factor in Eq. \eqref{eq:spwv} to be sufficiently large. When this weak value factor is large it will \emph{amplify} the linear response. Critically, the initial and final states for the weak values $S_w$ and $p_w$ can be strategically chosen to produce a large amplification factor. This is the essence of the technique used in weak value amplification \cite{Hosten2008,Dixon2009,Starling2010,Starling2010b,Turner2011,Hogan2011,Pfeifer2011,Zilberberg2011,Kedem2012,Puentes2012,Zhou2012,Egan2012,Gorodetski2012,Shomroni2013,Strubi2013,Xu2013,Viza2013,Zhou2013,Hayat2013,Magana2013,Jayaswal2014}.

For a tangible example of how this amplification works for estimating $\epsilon$, consider the measurement in Fig. \ref{crystal}(b). Averaging the position recorded at every pixel produces the centroid
\begin{align}\label{eq:condx}
  \int \! x\, P_\epsilon(x | \theta)\,\textrm{d}x &\approx \frac{\mean{x} + \epsilon (\mean{x^2}/\sigma^2) \text{Re}S_w}{1 + \epsilon (\mean{x}/\sigma^2) \text{Re}S_w}, \\
  &= \epsilon\, \text{Re}S_w. \nonumber
\end{align}
To compute Eq.~\eqref{eq:condx} we used the perturbed conditional probability $P_\epsilon(x|\theta) = P_\epsilon(x,\theta)/\int\! P_\epsilon(x,\theta)\textrm{d}x$ computed from Eq. \eqref{eq:Pex} as a function of the pixel position $x$, and a given postselection polarization angle $\theta$, as well as the Gaussian moments $\mean{x}=0$ and $\mean{x^2} = \sigma^2$ of the unperturbed beam profile.  Dividing the measured centroid by the (known) quantity $\text{Re}S_w$ allows us to determine the small parameter $\epsilon$.

Alternatively, if the CCD measures the Fourier plane as in Fig. \ref{crystal}(c), then each pixel corresponds to a transverse momentum.  Finding the centroid in this case produces
\begin{align}\label{eq:condp}
  \int p\, P_\epsilon(p | \theta)\,\textrm{d}p &\approx \frac{\mean{p} + 2 \epsilon \mean{p^2} \text{Im}S_w/\hbar}{1 + 2 \epsilon \mean{p} \text{Im}S_w/\hbar} \\
  &= \epsilon \frac{\hbar}{2\sigma^2}\text{Im}S_w, \nonumber
\end{align}
where we used Eq. \eqref{eq:Pep} and the Gaussian moments $\mean{p}=0$ and $\mean{p^2} = (\hbar/2\sigma)^2$ of the unperturbed beam profile.

\begin{figure}[t]
  \begin{center}
  \includegraphics[width=\columnwidth]{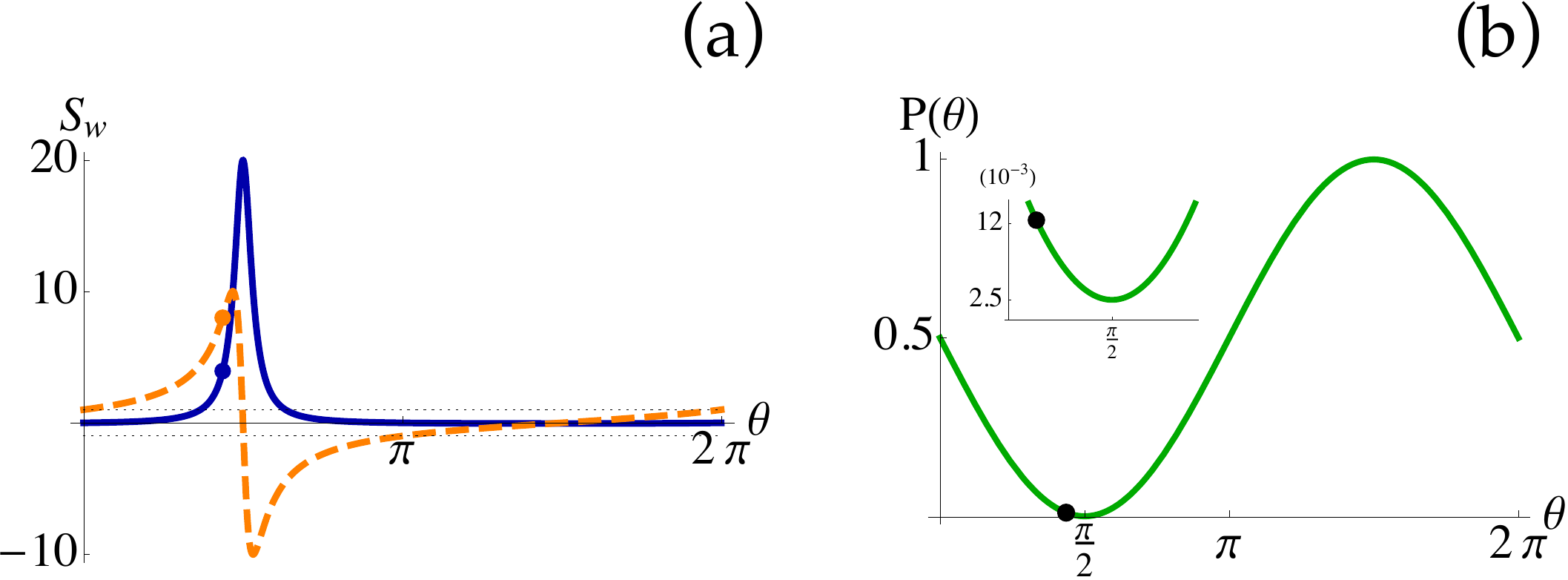}
  \end{center}
  \caption{(a) Real (dashed) and imaginary (solid) parts of the polarization weak value $S_w = \bra{f}\op{S}\ket{i}/\braket{f}{i}$, with initial state $\ket{i}$ given in Eq.~\eqref{eq:ipol} and shown in Fig.~\ref{fig:gauss1}, and final state $\ket{f}$ that depends on a varying angle $\theta$.  The eigenvalue bounds of $\pm 1$ are shown as dotted lines for reference, while the dots indicate the final state chosen in Eq.~\eqref{eq:fpol}. (b) The postselection probability $P(\theta)=\abs{\braket{f}{i}}^2$ as a function of $\theta$, showing how a large weak value corresponds to a small detection probability.  The inset shows the small probability region enlarged for clarity, while the dots similarly indicate the final state in Eq.~\eqref{eq:fpol}. }
  \label{fig:wv}
\end{figure}

The amplification occurs in each case because the factor $\text{Re}S_w$ in Eq.~\eqref{eq:condx} or $2\mean{p^2}\text{Im}S_w$ in Eq.~\eqref{eq:condp} can be made large by clever choices of polarization postselection.  For our example states [Eqs.~\eqref{eq:ipol} and \eqref{eq:fpol}], the polarization weak value is $S_w = \bra{f}\op{S}\ket{i}/\braket{f}{i} \approx 7.5 + 3.2 i$.  Notably, both the real and imaginary parts of the weak value in this case are larger than $1$, which is the maximum eigenvalue of $\op{S}$.  The plot in Fig.~\ref{fig:wv}(a) shows how the real and imaginary parts of the weak value vary with the choice of postselection angle $\theta$.

One cannot obtain such amplification to the sensitivity for free, however.  As the weak value factor $S_w$ becomes large, the detection probability necessarily decreases, as shown in Fig. \ref{fig:wv}(b).  Hence, the weak interaction approximation that assumes $2 \epsilon \text{Im}(S \otimes p)_w \ll \abs{\braket{f}{i}}^2 \abs{\braket{\psi_f}{\psi_i}}^2$ for each pixel will eventually break down and it will be necessary to include higher-order terms in $\epsilon$ that have been neglected, spoiling the linear response \cite{Geszti2010,Shikano2010,Cho2010,Shikano2011,Wu2011,Parks2011,Zhu2011,Koike2011,DiLorenzo2012,Dressel2012d,Nakamura2012,Susa2012,Pan2012,Dressel2012e,Wu2012,Kofman2012}.  Moreover, the resulting low detection rate (i.e., collected beam intensity) make it difficult to detect the signal, leading to longer collection times in order to overcome the noise floor.  Indeed, a careful analysis shows that the signal-to-noise ratio for determining $\epsilon$ within a fixed time duration remains constant as the amplification increases \cite{Starling2009,Feizpour2011,Jordan2014,Ferrie2014,Knee2014}---the signal gained by increasing the amplification factors in Eq.~\eqref{eq:condx} or \eqref{eq:condp} will exactly cancel the uncorrelated shot noise gained by decreasing the detection rate.  The scheme can also be sensitive to decoherence during the measurement \cite{Knee2013}.

Nevertheless, there are two distinct advantages to using this amplification technique: (1) the detector collects a fraction of the total beam power due to the postselection polarizer yet still shows similar sensitivity to optimal estimation methods \cite{Jordan2014,Knee2014,Pang2014}, and (2) the weakness of the measurement itself makes the amplification robust against certain types of additional technical noise (such as $1/f$ noise) \cite{Starling2009,Feizpour2011,Jordan2014,Ferrie2014,Knee2014}.  The former advantage allows less expensive equipment to be used, while simultaneously enabling the uncollected beam power to be redirected elsewhere for other purposes \cite{Starling2010,Dressel2013}.  The latter advantage allows one to amplify the signal without also amplifying certain types of unrelated (but common) technical noise backgrounds.  These two advantages combined are precisely what has permitted experiments such as \cite{Hosten2008,Dixon2009,Starling2010,Starling2010b,Turner2011,Hogan2011,Pfeifer2011,Zhou2012,Egan2012,Xu2013,Magana2013,Zhou2013,Jayaswal2014} to achieve such phenomenal precision with relatively modest laboratory equipment.

\subsection{Measurable complex value}
Since weak values are measurable complex quantities, they can be used to directly measure other normally inaccessible complex quantities in the quantum theory that can be expanded into sums and products of complex weak values, such as the geometric phase \cite{Sjoqvist2006,Kobayashi2010,Kobayashi2011}. Most notably, one can ``directly'' measure the quantum state itself using this technique \cite{Lundeen2011,Massar2011,Zilberberg2011,Lundeen2012,Salvail2013,Wu2013,Fischbach2012,Kobayashi2013,Malik2014}. Conventionally, a quantum state is determined through the indirect process of quantum tomography \cite{Altepeter2005}. Like its classical counterpart, quantum tomography involves making a series of projective measurements in different bases of a quantum state. This process is indirect in that it involves a time consuming postprocessing step where the density matrix of the state must be \emph{globally} reconstructed through a numerical search over the alternatives consistent with the measured projective slices. Propagating experimental error through this reconstruction step can be problematic, and the computation time can be prohibitive for determining high-dimensional quantum states, such as those of orbital angular momentum. 

We can bypass the need for such a global reconstruction step by expanding individual components of a quantum state directly in terms of measurable weak values. For a simple example, we determine the complex components of the initial polarization state $\ket{i}$ from Section~\ref{sec:measure}, as expanded in the weak measurement basis $\{\ket{H},\ket{V}\}$. This is accomplished by the insertion of the identity and multiplication by a strategically chosen constant factor $c = \braket{D}{H}/\braket{D}{i} = \braket{D}{V}/\braket{D}{i}$, where the postselection state $\ket{D}$ is unbiased with respect to both $\ket{H}$ and $\ket{V}$.  With this clever choice the scaled state has the form
\begin{align}
  c \ket{i} = \underbrace{\frac{\braket{D}{H}\braket{H}{i}}{\braket{D}{i}}}_{H_w}\ket{H}+ \underbrace{\frac{\braket{D}{V}\braket{V}{i}}{\braket{D}{i}}}_{V_w}\ket{V}.
\end{align}
That is, each complex component of the scaled state $c \ket{i}$ can be directly measured as a complex first order weak value.  After determining these complex components experimentally, the state can be subsequently renormalized to eliminate the constant $c$ up to a global phase.

Furthermore, we can write the projections as $\ket{H}\bra{H} = (\op{1} + \op{S})/2$ and $\ket{V}\bra{V} = (\op{1} - \op{S})/2$, so we can rewrite the required weak values $H_w = (1 + S_w)/2$ and $V_w = (1 - S_w)/2$ in terms of the single polarization weak value $S_w$. We showed earlier how to isolate and measure both the real and imaginary parts of this polarization weak value.  Thus, we can completely determine the state $\ket{i}$ after the polarization weak value $S_w$ has been measured using the special postselection $\ket{D}$. 

The primary benefit of this direct state estimation approach is that minimal postprocessing (and thus minimal experimental error propagation) is required to reconstruct individual state components from the experimental data. The real and imaginary parts of each pure state component in a desired basis directly appear in the linear response of a measurement device up to appropriate scaling factors.  Mixed states can also be measured in a similar way by scanning the postselection across a mutually unbiased basis, which will determine the Dirac distribution for the state instead \cite{Lundeen2012,Salvail2013,Lundeen2014}; this distribution is related to the density matrix via a Fourier transform. 

The downside of this approach is that the denominator $\braket{D}{i}$ in the constant $c$ cannot become too small or the linear approximation used to measure $S_w$ will break down \cite{Haapasalo2011}, causing estimation errors \cite{Maccone2014}. This restriction limits the generality of the technique for faithfully estimating a truly unknown state. Furthermore, improperly calibrating the weak interaction can introduce unitary errors or produce additional decoherence that does not appear in projective tomography techniques. Nevertheless, the direct measurement technique can be useful for determining the components of most states.

\subsection{Conditioned average}
\begin{figure}[t]
  \begin{center}
  \includegraphics[width=\columnwidth]{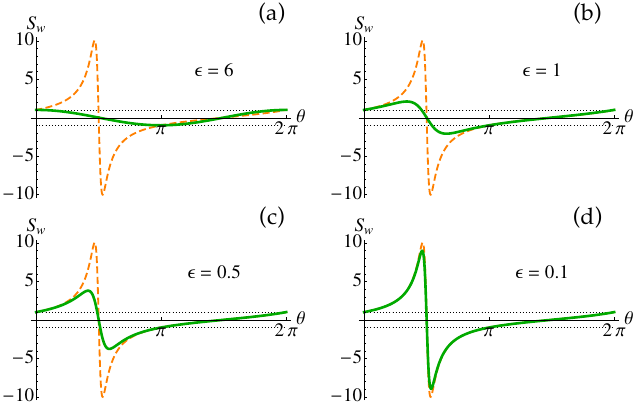}
  \end{center}
  \caption{Conditioned average \eqref{eq:cavwv} of generalized polarization eigenvalues $x/\epsilon$ for various values of the crystal length $\epsilon$, using the beam profile illustrated in Figure~\ref{fig:gauss1}.  For large $\epsilon$ the average is a classical conditioned average constrained to the eigenvalue range (dotted lines).  For sufficiently small $\epsilon$, however, the conditioned average (solid lines) approximates the real part (dashed lines) of the polarization weak value in Fig.~\ref{fig:wv}.}
  \label{fig:cav1}
\end{figure}

As our final example of the utility of weak values, we show that the real part of a weak value can be interpreted as a form of conditioned average associated with an observable.  To show this we first consider how each pixel records polarization information in the absence of postselection.  After summing over all complementary postselections $\ket{f}$ in the perturbed probability $P_\epsilon(x,f)$ in Eq.~\eqref{eq:pepsilon}, we can express the total perturbed pixel probability as
\begin{align}\label{eq:totpx}
  P_\epsilon(x) &= \sum_f \abs{\bra{f}\bra{x}e^{-i\epsilon\op{S}\otimes\op{p}/\hbar}\ket{i}\ket{\psi_i}}^2 = \bra{i} \op{P}_x \ket{i},
\end{align}
in terms of a probability operator 
\begin{align}\label{eq:totpxop}
  \op{P}_x &= \abs{\braket{x - \epsilon}{\psi_i}}^2\,\ket{H}\bra{H} + \abs{\braket{x + \epsilon}{\psi_i}}^2\,\ket{V}\bra{V}, \nonumber \\
 &= \abs{\braket{x-\epsilon \op{S}}{\psi_i}}^2.
\end{align}
The second line is a formal way of writing the probability operator more compactly in terms of the spectral representation of $\hat S$. This formal expression also supports the intuition that $\hat P_x$ indicates that the crystal interaction shifts the initial profile $\abs{\braket{x}{\psi_i}}^2$ of the beam by an amount that depends on the polarization.

An experimenter can then \emph{assign} a value of $(x/\epsilon)$ to each pixel $x$ and average those values over the perturbed profile in Eq.~\eqref{eq:totpx} to obtain the average polarization
\begin{align}\label{eq:totsav}
  \int \frac{x}{\epsilon}\, P_\epsilon(x)\,\mathrm{d}x &= \bra{i}\op{S}\ket{i}
\end{align}
for any preparation state $\ket{i}$.  The values $(x/\epsilon)$ assigned to each pixel act as \emph{generalized eigenvalues} for the polarization operator $\op{S}$ \cite{Dressel2010,Dressel2012,Dressel2012b}.  An experimenter must assign these values in place of the standard polarization eigenvalues of $\pm 1$ because the pixels are only weakly correlated with the polarization.  Although the values $(x/\epsilon)$ generally lie well outside the eigenvalue range of $\op{S}$, their experimental average in Eq.~\eqref{eq:totsav} always produces a sensible average polarization.

The state independence of this procedure can be emphasized by noting that the assignment of the generalized eigenvalues $(x/\epsilon)$ formally produces an operator identity,
\begin{align}\label{eq:cvs}
  \int \frac{x}{\epsilon}\, \op{P}_x\, \mathrm{d}x &= \op{S}
\end{align}
in terms of the probability operators $\op{P}_x$ in \eqref{eq:totpxop} that correspond to each measured pixel.  This identity guarantees that the experimenter can faithfully reconstruct information about the observable $\op{S}$ for any unknown state by properly weighting the probabilities for measuring each CCD pixel.  In the special case of a projective measurement, the probability operators will be the spectral projections for $\op{S}$ and the assigned values will be the eigenvalues of $\op{S}$, which makes Eq.~\eqref{eq:cvs} a natural generalization of the spectral expansion of $\op{S}$ to a generalized measuring apparatus.

It is worth noting that since there are more pixels than polarization eigenvalues, one can form an operator identity such as Eq.~\eqref{eq:cvs} in many different ways by assigning different values $\alpha(x)$ to the pixel probabilities.  In such a case, the information redundancy in the pixel probabilities gives the freedom to choose appropriate values that statistically converge more rapidly to the desired mean \cite{Dressel2010,Dressel2012,Dressel2012b}.  For our purposes here, however, we use the simplest generic choice $\alpha(x) = x/\epsilon$.

Including the effect of the postselection polarizer $\ket{f}$ changes this general result.  The added polarizer conditions the total pixel probability of Eq.~\eqref{eq:totpx}.  After assigning the same generalized polarization eigenvalues $x/\epsilon$ to each pixel and averaging these values over the conditioned profile, an experimenter will find the \emph{conditioned average}
\begin{align}\label{eq:cavwv}
  \int \frac{x}{\epsilon}\, P_\epsilon(x|f)\,\mathrm{d}x = \text{Re}\frac{\bra{f}\op{S}\ket{i}}{\braket{f}{i}} + O(\epsilon^2).
\end{align}
As shown in Eq.~\eqref{eq:condx} this conditioned average of generalized polarization eigenvalues approximates the real part of a weak value for small $\epsilon$ in an experimentally meaningful way. 

Importantly, even when $\epsilon$ is not small the full conditioned average of generalized eigenvalues \eqref{eq:cavwv} will smoothly interpolate between the weak value approximation and a classical conditioned average of polarization. In Fig.~\ref{fig:cav1} we illustrate this interpolation for different values of $\epsilon$. This smooth correspondence is essential for associating the experimental average Eq.~\eqref{eq:cavwv} to the polarization $\op{S}$ in any meaningful way. Indeed, we have shown \cite{Dressel2012d,Dressel2012e} that this interpolation exactly describes how the initial polarization state \emph{decoheres} into a classical polarization state with increasing measurement strength. Moreover, this technique of constructing conditioned averages of generalized eigenvalues works quite generally for other detectors \cite{Pryde2005,Romito2008,Kedem2010a,Dressel2011,Goggin2011,Dressel2012c,Weston2013,Zilberberg2013,Silva2014} and produces similar interpolations between a classical conditioned average and the real part of a weak value.

The link between weak values and conditioned averages has been used to address several quantum paradoxes, such as Hardy's paradox \cite{Aharonov2002,Lundeen2009,Yokota2009} and the three-box paradox \cite{Resch2004}.  Anomalously large weak values provide a measurable window into the inner workings of these paradoxes by indicating when quantum observables cannot be understood in any classical way as properties related to their eigenvalues.  Similarly, anomalously large weak values have been linked to violations of generalized Leggett-Garg inequalities \cite{Williams2008,Palacios-Laloy2010,Goggin2011,Dressel2011,Suzuki2012,Emary2014,Groen2013} that indicate nonclassical (or invasive) behavior in measurement sequences.  This link has also been exploited to provide an experimental method for determining physically meaningful conditioned quantities, such as group velocities in optical fibers \cite{Brunner2004}, or the momentum-disturbance relationships for a two-slit interferometer \cite{Mir2007}.  

A particularly notable experimental demonstration of the connection between weak values and physically meaningful conditioned averages is the measurement of the locally averaged momentum streamlines $p_B(x)$ passing through a two-slit interferometer performed by \citet{Kocsis2011} using the weak value identity
\begin{align}\label{eq:bohm}
  \text{Re}\frac{\bra{x}\op{p}\ket{\psi_i}}{\braket{x}{\psi_i}} = \partial_x \Phi(x) = p_B(x),
\end{align}
where $\braket{x}{\psi_i} = |\braket{x}{\psi_i}| \exp[i \Phi(x)/\hbar]$ is the polar decomposition of the initial transverse profile.  This phase gradient has appeared historically in Madelung's hydrodynamic approach to quantum mechanics \cite{Madelung1926,Madelung1927}, Bohm's causal model \cite{Bohm1952a,Bohm1952b,Wiseman2007,Traversa2013}, the momentum part of the local energy-momentum tensor \cite{Hiley2012}, and even the Poynting vector field of classical electrodynamics \cite{Bliokh2013,Dressel2014}.  Importantly, the weak value connection provides this quantity with an experimentally meaningful definition as a weakly measured conditioned average.

\section{Conclusions}

In this Colloquium we explored how the quantum weak value naturally appears in laboratory situations.  We operationally defined weak values as complex parameters that completely characterize the relative corrections to detection probabilities that are caused by an intermediate interaction.  When the interaction is sufficiently weak, these relative corrections can be well approximated by first order weak values.

Using an optical example of a polarized beam passing through a birefringent crystal, we showed how to use a product interaction to isolate and measure both the real and imaginary parts of first order weak values.  This example allowed us to discuss three distinct roles that the first order weak value has played in recent experiments.

First, we showed how a large weak value can be used to amplify a signal used to sensitively estimate an unknown interaction parameter in the (linear) weak interaction regime. Although the signal-to-noise ratio remains constant from this amplification due to a corresponding reduction in detection probability, the technique allows one to amplify the signal above other technical noise backgrounds using fairly modest laboratory equipment.

Second, we showed that since the first order weak value is a measurable complex parameter, it can be used to experimentally determine other complex theoretical quantities.  Notably, we showed how the components of a pure quantum state may be directly determined up to a global phase by measuring carefully chosen weak values.

Third, we discussed the relationship between the real part of a first order weak value and a conditioned average for an observable.  By conditionally averaging generalized eigenvalues for the observable, we showed that one obtains an average that smoothly interpolates between a classical conditioned average and a weak value as the interaction strength changes.

We have emphasized the generality of the quantum weak value as a tool for describing experiments.  Because of this generality, we anticipate that many more applications of the weak value will be found in time.  We hope this Colloquium will encourage further exploration along these lines.

\begin{acknowledgments}
\emph{Acknowledgments}.---JD and ANJ acknowledge support from the National Science Foundation under Grant No. DMR-0844899, and the US Army Research Office under Grant No. W911NF-09-0-01417. MM, FMM, and RWB acknowledge support from the US DARPA InPho program. FMM and RWB acknowledge support from the Canada Excellence Research Chairs Program. MM acknowledges support from the European Commission through a Marie Curie fellowship. The authors thank Jonathan Leach for helpful discussions.
\end{acknowledgments}

%
\end{document}